\documentstyle[amssymb,aps,epsfig,multicol]{revtex}

\begin{document}

\title{Effect of External Noise Correlation in Optical Coherence Resonance}

\author{J.M. Buld\'u,$^1$ J. Garc\'{\i}a-Ojalvo,$^1$ Claudio R. Mirasso,$^2$
M.C. Torrent,$^1$ and J.M. Sancho$^3$}

\address{$^1$Departament de F\'{\i}sica i Enginyeria Nuclear, Universitat
Polit\`ecnica de Catalunya, Colom 11, E-08222 Terrassa, Spain\\
$^2$Departament de F\'{\i}sica, Universitat de les Illes Balears,
E-07071 Palma de Mallorca, Spain\\
$^3$Departament d'Estructura i
Constituents de la Mat\`eria, Universitat de Barcelona, Diagonal
647, E-08028 Barcelona, Spain}

\maketitle

\begin{abstract}
Coherence resonance occurring in semiconductor lasers with optical
feedback is studied via the Lang-Kobayashi model with external
non-white noise in the pumping current. The temporal correlation and the
amplitude of the noise have a highly relevant influence in
the system, leading to an optimal coherent response for suitable
values of both the noise amplitude and correlation time. This
phenomenon is quantitatively characterized by means of several
statistical measures.

\end{abstract}


\begin{multicols}{2}
{

Despite random fluctuations usually constitute a source of
disorder in dynamical systems, many examples exist in which they
lead instead to an increase of order in the system behavior. Among
these examples, {\em stochastic resonance} stands out for its
implications in many different areas of science
\cite{wiesenfeld95}. In the conventional situation, stochastic
resonance consists of an optimization due to noise of the response
of a nonlinear system to a weak periodic driving
\cite{gammaitoni98}. But even in the absence of external periodic
forcing, noise can be helpful in sustaining a coherent
oscillatory response in the system, provided the operation point is
close to a limit cycle \cite{GDNH93} or within an excitable regime
\cite{PK97}. This phenomenon has been called {\em coherence resonance}
(CR), and has been recently found also in bistable \cite{lutz} and
chaotic \cite{claudio} systems.

One of the earliest and most influential experimental observations
of stochastic resonance was made in a laser system \cite{raj}.
Similarly, optical systems have also provided in recent years clear-cut
examples of excitable behavior, including observations in
semiconductor lasers subject to optical feedback \cite{jorge97,josep99},
lasers with saturable absorber \cite{stefano},
passive nonlinear ring cavities \cite{harrison},
lasers with injected signal \cite{coullet98},
and self-pulsing lasers \cite{DK99}.
Following these studies, optical CR was
predicted theoretically in the self-pulsing laser \cite{DKL99} and
observed experimentally in the semiconductor laser with optical feedback
\cite{tredicce}.
In this latter case, noise is added to the driving current of the laser,
and gives rise to a pulsed behavior in the system, in the form of
sudden drop-outs in the evolution of the light intensity. The regularity
of the drop-out series initially increases with increasing fluctuations,
and peaks for an optimal amount of noise.
The present Communication is devoted to the theoretical
modeling of this situation, making use of a rate equation system 
including a delay term, i.e. the well-know Lang-Kobayashi (LK) model
\cite{lang} generalized to take into account the insertion of external
noise into the system through the laser's pumping current.

Our results show that a white-noise assumption is not
adequate to account for the observed resonant behavior. In fact, such
a supposition is not realistic, due to the fast time scales in which this
system evolves ($\sim$ tens of ps), smaller than or on the order of the
characteristic time scales of the fastest fluctuations that can be
experimentally introduced, which are restricted by 0 limitations of the
electronics involved ($\sim\,$GHz). Following these considerations, we have
considered a {\em time-correlated} external noise, and found that coherence
is maximal not only for an optimal noise amplitude, but also for an optimal
noise correlation time. In what follows, such a double coherence resonance
is described in detail and characterized by suitable statistical measures.

The LK model describes the temporal evolution of the slowly varying
complex envelope of the electric field $E(t)$ inside the laser and the
excess carrier number $N(t)$, considering only one longitudinal mode of
the solitary laser, and one single reflection from the external feedback
mirror (i.e. multiple reflections are neglected, which is valid for not
too large reflectivities). In dimensionless form the model reads
\cite{josep99,lang}:
\begin{eqnarray}
\frac{d E}{d t} & = & \frac{1+i\alpha}{2}(G(E,N)-\gamma)\,E(t)
\nonumber
\\
& &\qquad\qquad+\kappa e^{-i\omega\tau_f} E(t-\tau_f)
+\sqrt{2\beta N} \zeta(t)
\label{eq:LK-model}
\\
\frac{d N}{d t} & = & \gamma_e\left\{C[1+\xi(t)]N_{\rm th}-N(t)\right\}
-G(E,N)\,|E(t)|^2\,,
\nonumber
\end{eqnarray}
where $\gamma$ and $\gamma_e$ are the inverse lifetimes of photons and
carriers, respectively, $C$ is the pumping rate (directly related to
the driving current; $C=1$ is the solitary-laser threshold),
$\alpha$ is the linewidth enhancement factor,
and $\omega$ is the solitary-laser frequency. The
last term in the electric-field equation represents spontaneous emission
fluctuations, with $\zeta(t)$ a Gaussian white noise of zero mean and
unity intensity, and
$\beta$ measuring the internal noise strength.
The material-gain function $G(E,N)$ is given by
\begin{equation}
G(E,N) = \frac{g(N(t)-N_0)}{1+s|E(t)|^2}\,,
\label{eq:ganancia}
\end{equation}
where $g$ is the differential gain coefficient and $s$ the saturation
coefficient. The threshold carrier number is $N_{\rm th}=\gamma/g+N_0$.
The optical feedback is described by two parameters: the
feedback strength $\kappa$ and the external round--trip time $\tau_f$.
Finally, the external noise is represented by the term $\xi(t)$,
which according to the discussion made above, is taken to be a
time-correlated noise of the Ornstein-Uhlenbeck type, gaussianly distributed
with zero mean and correlation
\begin{equation}
\langle\xi(t)\xi(t')\rangle=
\frac{D}{\tau_c}\,e^{-(t-t')/\tau_c}
\label{eq:oun}
\end{equation}
This external noise is characterized by two parameters, its intensity
$D$ and its correlation time $\tau_c$. The variance of the noise is given
by $D/\tau_c$, and hence we will measure its amplitude as
$\sigma=\sqrt{D/\tau_c}$.

The LK model with no external noise has been
profusely used in the past to model the dynamics of semiconductor lasers
subject to optical feedback. In particular, it satisfactorily describes
the appearance of drop-outs when the system is perturbed with electrical
pulses over a threshold value, with the feature that the shape of the
generated (inverted) pulses is basically independent of the perturbation
\cite{josep99}. This behavior, which is characteristic of excitable systems
and appears when the laser is operated
close to the solitary laser threshold, agrees with experimental observations
\cite{jorge97}. For larger pumping rates (but still close to the laser
threshold) the intensity dropouts appear spontaneously, i.e. no external
excitation is needed to produce them.
Figure \ref{fig:int-pha}
displays an example of this behavior, in terms of the evolution of both
the intensity $I(t)$ and phase difference between consecutive round-trips
$\eta(t)=\phi(t)-\phi(t-\tau_f)$, with $E(t)=\sqrt{I}\exp(i\phi)$.
The occurrence of pulses in the form of power drop-outs in the intensity
time series can be clearly identified, and they are seen to correspond
with well-defined pulses in the electric-field phase difference.
\begin{figure}[htb]
\epsfig{file=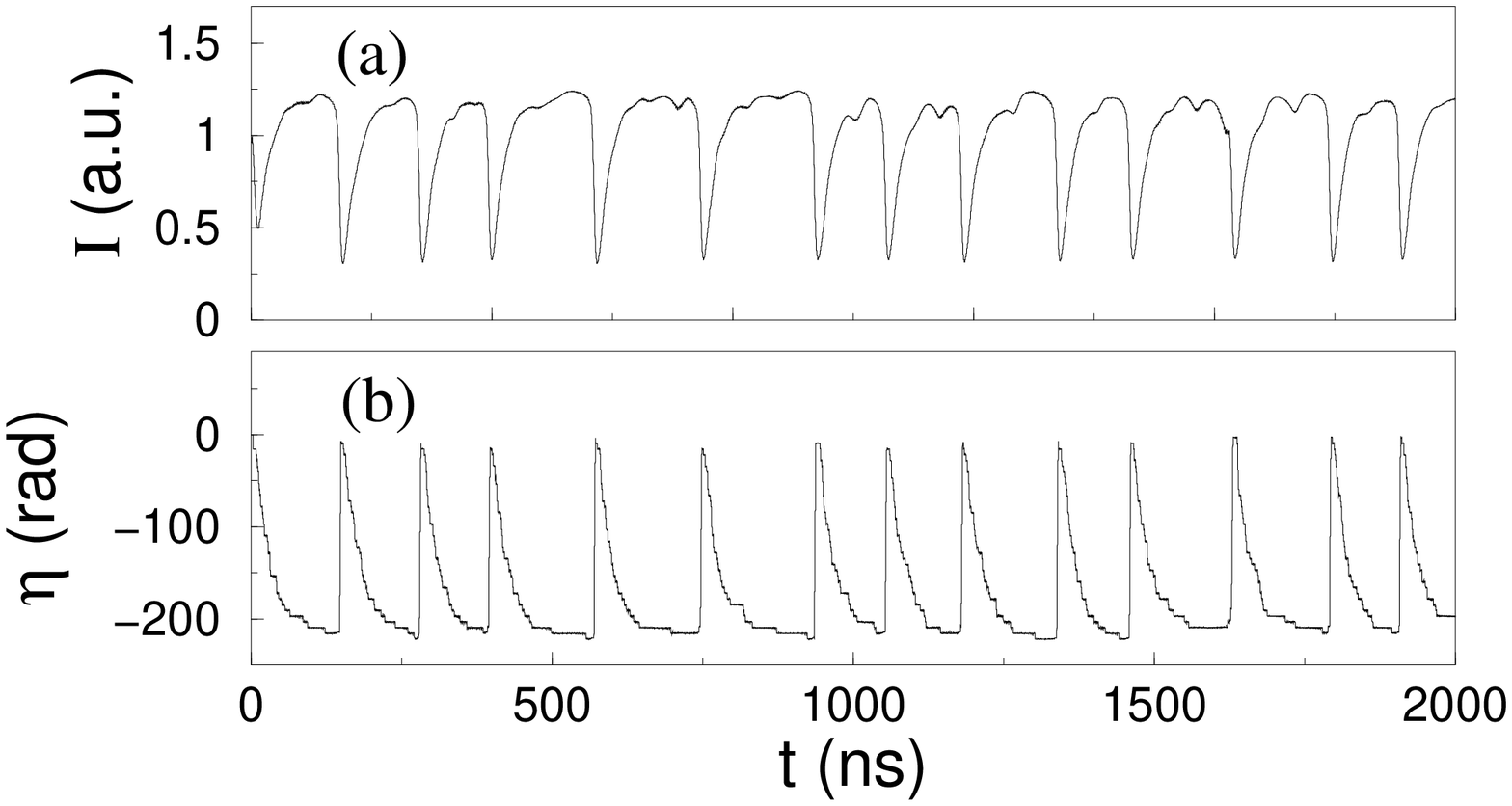,width=70mm}
\caption{
Time evolution exhibiting drop-outs of the intensity $I$ and the corresponding
pulses in the phase difference $\eta$. Parameters of the LK model are:
$C=1.03$,
$\gamma_e=6\cdot 10^{-4}$~ps$^{-1}$,
$\gamma=0.158$~ps$^{-1}$,
$g=2.79\cdot 10^{-9}$~ps$^{-1}$,
$s=3\cdot 10^{-7}$,
$\alpha=3.5$,
$N_0=1.51\cdot 10^{8}$,
$\beta=5\cdot 10^{-10}$~ps$^{-1}$,
$\kappa=0.025$~ps$^{-1}$,
$\tau_f=2.4$~ns,
$\omega\tau_f=2$,
and $D=0$.
}
\label{fig:int-pha}
\end{figure}
It should be noted that the intensity time trace shown in Fig.
\ref{fig:int-pha}(a) has been filtered to $100$~MHz,
in order to mimic the bandwidth effect of typical photodetectors. For
large enough filtering bandwidth (or for no filtering at all), the
corresponding evolution takes the form of ultrashort intensity pulses
(with durations on the order of tens of ps) \cite{fast}, whose envelope
exhibits the low-frequency drop-outs shown in Fig. \ref{fig:int-pha}(a).

Since the internal spontaneous-emission noise $\zeta(t)$
cannot be experimentally controlled, we now turn our attention to the
effect of the external noise $\eta(t)$ upon the system. This effect
can be understood by examining
the mechanism behind the above-mentioned power drop-outs, which is well
understood in the framework of the LK model (i.e. under the assumption
of single-mode operation of the laser).
This model exhibits multiple
coexisting fixed points, which appear in pairs of solutions called modes
and antimodes. The antimodes are saddle points, and most of the modes
are also unstable due to a Hopf bifurcation \cite{lsa}. However, at least
one of the modes (the one with maximum power) is stable. In this complex
phase-space landscape, a large enough fluctuation may be able to take
the system away from the basin of attraction of the stable fixed point
and, upon collision with a neighboring antimode, produce a sudden increase
in the phase difference [see Fig. \ref{fig:int-pha}(b)] which corresponds
to a power drop-out. The corresponding escape time, also called
activation time $t_a$, is a random variable whose average decreases with
the intensity of the external noise according to Kramers'
law \cite{kramers}. Following the drop-out, a build-up process begins in
which the system undergoes a chaotic itinerancy around the Hopf-unstable
modes, jumping consecutively from one to the next while being drifted
back towards the stable maximum-gain mode \cite{sano}. For small intensities
of the external noise, the excursion time $t_e$ required by this process is
basically independent of noise, and has the role of a refractory time during
which no drop-outs can be induced. As noise intensity increases the escape
events become more frequent, reducing the standard deviation of the
interspike intervals accordingly.
A minimum of variability occurs for an optimal amount of noise when the
drop-out separation is of the order of $t_e$. Beyond that point, noise
intensity is large enough to produce escapes before the build-up process
is finished (i.e. before the stable mode is reached), which leads to an
irregular series of pulses. This sequence of events is depicted in Fig.
\ref{fig:ev-amp}, which shows three time traces of the phase difference
$\eta(t)$ for increasing amplitudes of the external noise, keeping its
correlation time constant. In this case, the semiconductor laser
is biased at $1\%$ above the solitary-laser threshold, a situation for
which the system is stable in the absence of external noise. A small
amount of noise produces infrequent drop-outs [Fig. \ref{fig:ev-amp}(a)],
which become more numerous and regular as the noise amplitude increases
[Fig. \ref{fig:ev-amp}(b)].
For large noise strengths the pulses become increasingly irregular, both
in separation and in amplitude
[Fig. \ref{fig:ev-amp}(c)].
Hence, an optimal amplitude of the external noise exists for which the
coherence of the pulsed output of the laser is optimal.

In order to quantitatively characterize this effect, we compute the
standard deviations $R_\theta$ and $R_\mu$ of the normalized drop-out
separation $\theta=T/\langle T \rangle$ and normalized drop-out
amplitude $\mu=A/\langle A \rangle$, respectively \cite{tredicce},
where the statistical averages are performed over both time and different
realizations of the noise.
Numerically, to detect a drop-out we study the behavior of the phase
difference $\eta(t)$, since its variation is smoother than that of the
intensity, and easier to characterize. A drop-out occurrence is recorded
when the phase difference $\eta$ suddenly increases in a fixed large amount
(e.g. $12\pi$ in our case) \cite{josep99}. The drop-out amplitude $A$
is a measure of the total height of the phase pulse.

\begin{figure}[htb]
\begin{center}
\epsfig{file=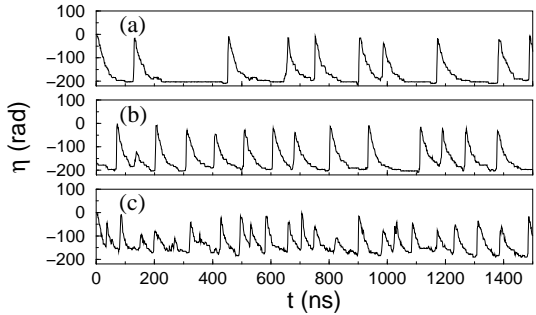,width=7.0cm}
\caption{
Temporal behavior of the phase difference $\eta$ for increasing noise
amplitude:
(a) $\sigma=7.36\cdot 10^{-2}$,
(b) $\sigma=9.35\cdot 10^{-2}$,
and (c) $\sigma=1.60\cdot 10^{-1}$.
In all cases we have kept the correlation time constant, $\tau_c=24$~ps.
Other parameters are those of Fig. \protect\ref{fig:int-pha}, except
$C=1.01$.
}
\label{fig:ev-amp}
\end{center}
\end{figure}

\begin{figure}[htb]
\centerline{
\epsfig{file=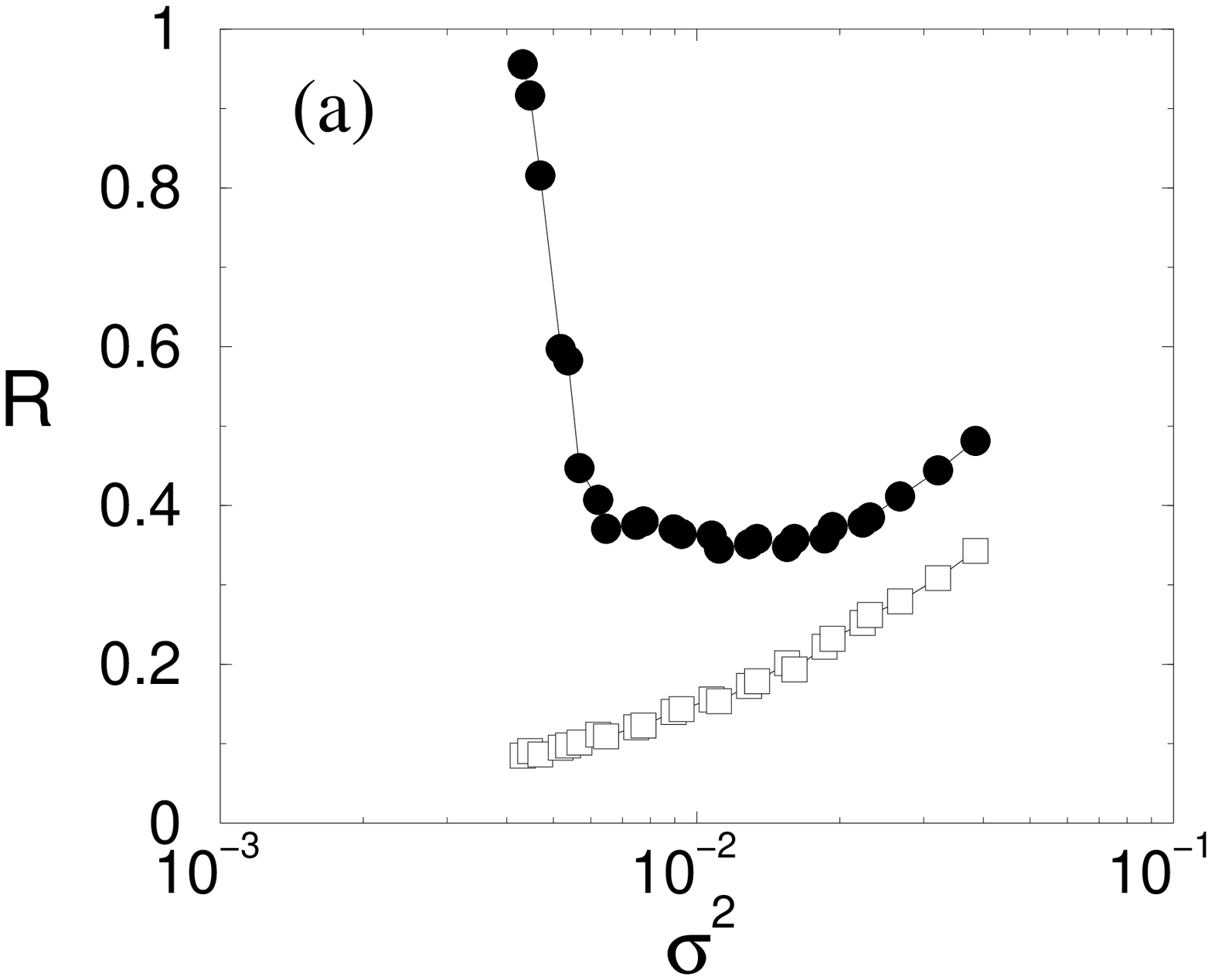,width=4.0cm}
\epsfig{file=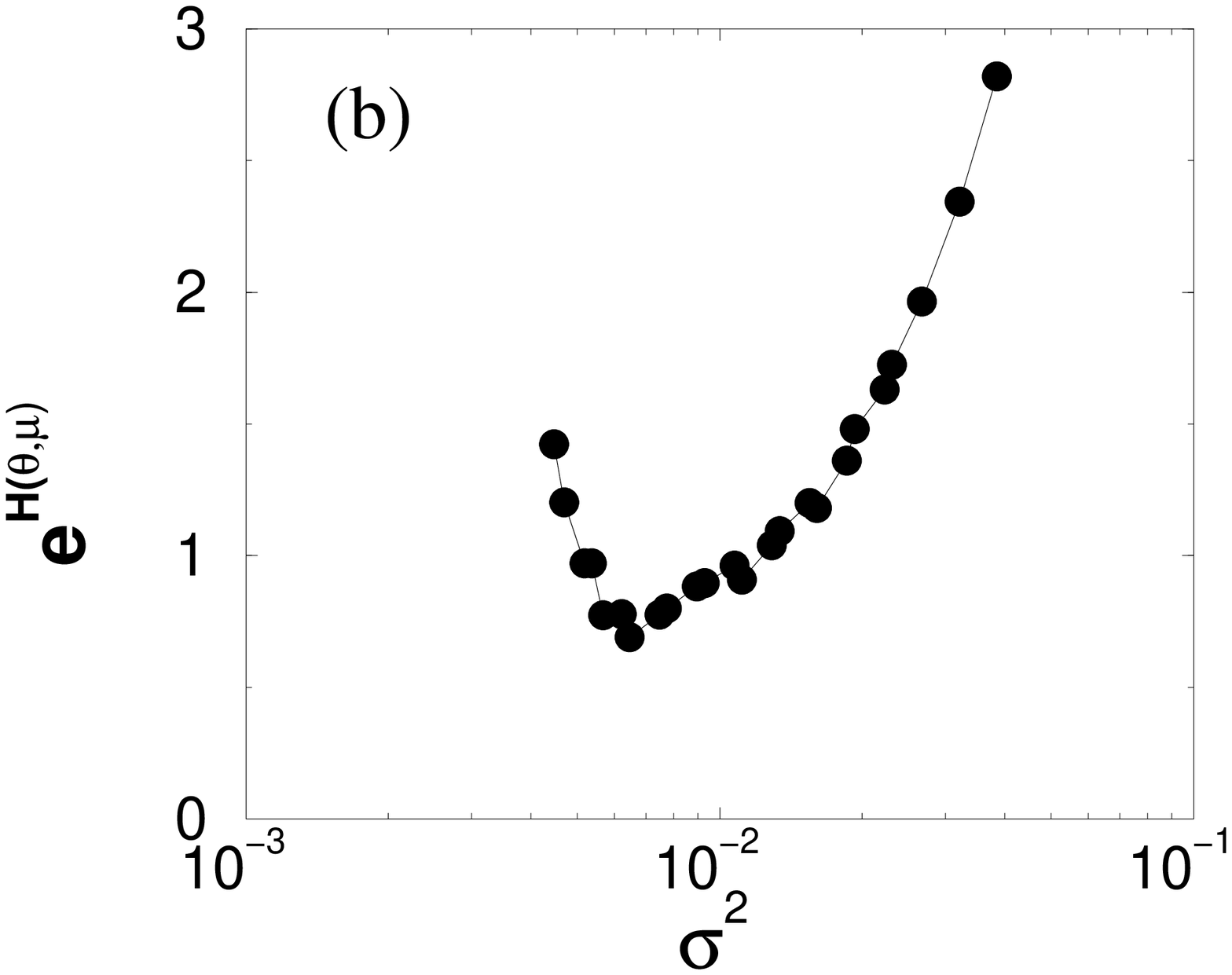,width=4.0cm}
}
\caption{
Statistical characterization of the noise-amplitude coherence resonance.
(a) Standard deviations $R_\theta$ (full circles) and $R_\mu$ (empty squares),
and (b) joint entropy $H(\theta,\mu)$ as a function of the external noise
amplitude. The time correlation of the noise is fixed to $\tau_c=24$~ps.
Other parameters are those of Fig. \protect\ref{fig:ev-amp}.
}
\label{fig:R-ent-amp}
\end{figure}

We have computed the standard deviations $R_\theta$ and $R_\mu$ for
increasing noise amplitude, averaging up to 20,000 drop-outs in each
measure. The result is
plotted in Fig. \ref{fig:R-ent-amp}(a), and confirms the qualitative
conclusions that have been drawn above from Fig. \ref{fig:ev-amp}, at
least as far as the variability of the pulse separation, $R_\theta$,
is concerned. This quantity is a non-monotonic function of the noise
amplitude, being minimal for an optimal amount of noise. The irregularity
of the drop-out amplitudes, on the other hand, increases monotonically 
with noise. This result coincides with experimental observations
\cite{tredicce}, and reflects the fact that the frequency with which noise
breaks up the build-up process increases steadily with the amount of noise
added.

In order to take into account {\em both} the drop-out separation and
amplitude simultaneously in the determination
of the signal's regularity, it is useful to define a joint entropy
$H(\theta,\mu)$ of the two quantities, where $H=-\sum P \log P$, with $P$
the joint probability density of the two random variables \cite{badii}.
For the case of two Gaussian independent random variables the following
relation holds \cite{tredicce}: $\exp(H(\theta,\mu))=2\pi e R_\theta R_\mu$.
We will assume that this result is approximately valid in our case, and
compute the joint entropy accordingly. The result is given in Fig.
\ref{fig:R-ent-amp}(b), which
shows again a maximum regularity of the drop-out series for an optimal
noise amplitude, this time taking into account both the pulse separation
and amplitude. In fact, the minimum is in this case more clearly defined.

We note that all results presented so far have been computed for a fixed
correlation time of the noise $\tau_c=24$~ps, on the order of the fast
time scale of the deterministic dynamics. In fact, as the white-noise
limit is approached the amount of noise necessary to obtain similar
effects climbs up to unreasonably high values. The reason is that the
carrier dynamics acts as a frequency filter for the external noise [see
equation for $N(t)$ in (\ref{eq:LK-model})], which also prevents the system
from responding to high-frequency modulations of the pump current. Therefore,
most of the power of a white noise has no effect upon the system dynamics,
and the noise intensity needs to be very large in order to have a noticeable
influence (a similar effect has been observed in periodic-modulation
studies \cite{sukow00}).
In the opposite frequency limit a similar situation occurs:
for low-frequency forcing the carrier dynamics has enough time to follow the
modulation, and the system responds simply with a modulated output.
Only for intermediate frequencies will the external forcing be able to
influence the drop-out statistics and enhance the coherent response of
the system. In order to verify this conjecture, we now fix the amplitude
$\sigma=\sqrt{D/\tau_c}$ of the external noise and analyze the behavior
of the system for an increasing correlation time of the Ornstein-Uhlenbeck
noise defined by Eq. (\ref{eq:oun}). The result is shown in Fig.
\ref{fig:ev-tcorr} for
three different values of $\tau_c$. It can be seen that the regularity
of the pulsed time series is maximal for intermediate values of the
noise correlation time.

We quantify again the qualitative observation made in the previous
paragraph by computing the standard deviations $R_\theta$ and $R_\mu$,
and the joint entropy $H(\theta,\mu)$ of the normalized drop-out separation
and amplitude. The results, shown in Fig. (\ref{fig:R-ent-tcorr}),
exhibit the same behavior as in the case of an increasing noise amplitude.
Coherence of the pulsed behavior is in this case maximal for a correlation
time of the noise $\tau_c\sim 30$~ps.
This behavior can be interpreted as a resonance with the fast deterministic
dynamics of the system. A similar resonance has been recently observed
experimentally in a chemical excitable medium \cite{irene}.

\begin{figure}[htb]
\epsfig{file=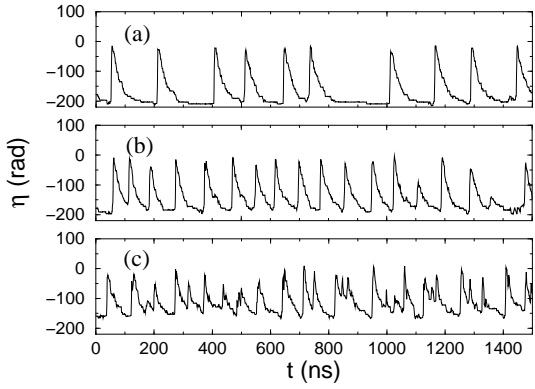,width=7cm}
\caption{
Temporal behavior of the phase difference $\eta$ for increasing noise
correlation time:
(a) $\tau_c=15.8$~ps,
(b) $\tau_c=57.6$~ps,
and (c) $\tau_c=153.2$~ps.
In all cases we have kept the noise amplitude constant, $\sigma=0.079$.
Other parameters are those of Fig. \protect\ref{fig:ev-amp}.
}
\label{fig:ev-tcorr}
\end{figure}

\begin{figure}[htb]
\centerline{
\epsfig{file=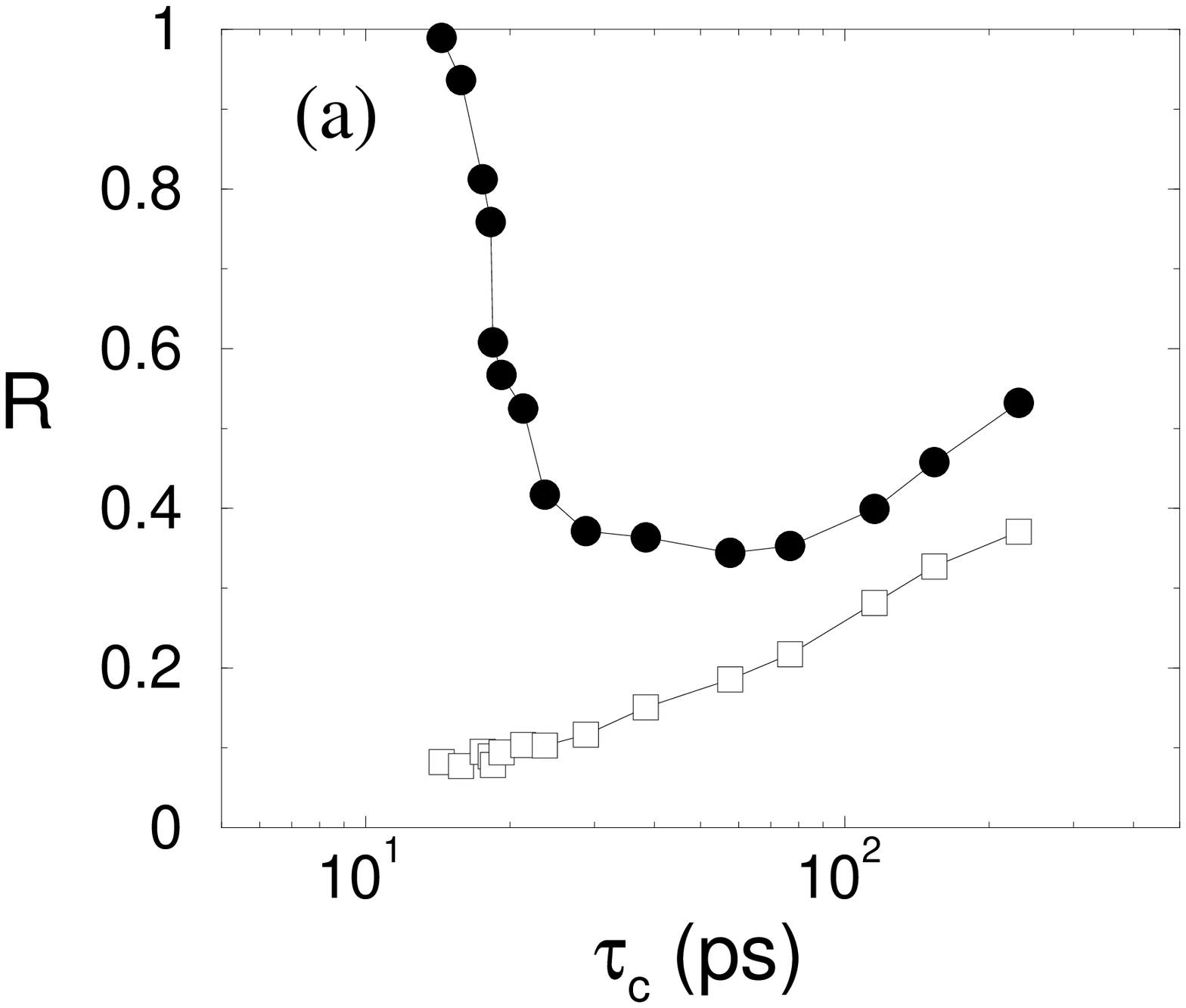,width=4cm}
\hskip2mm
\epsfig{file=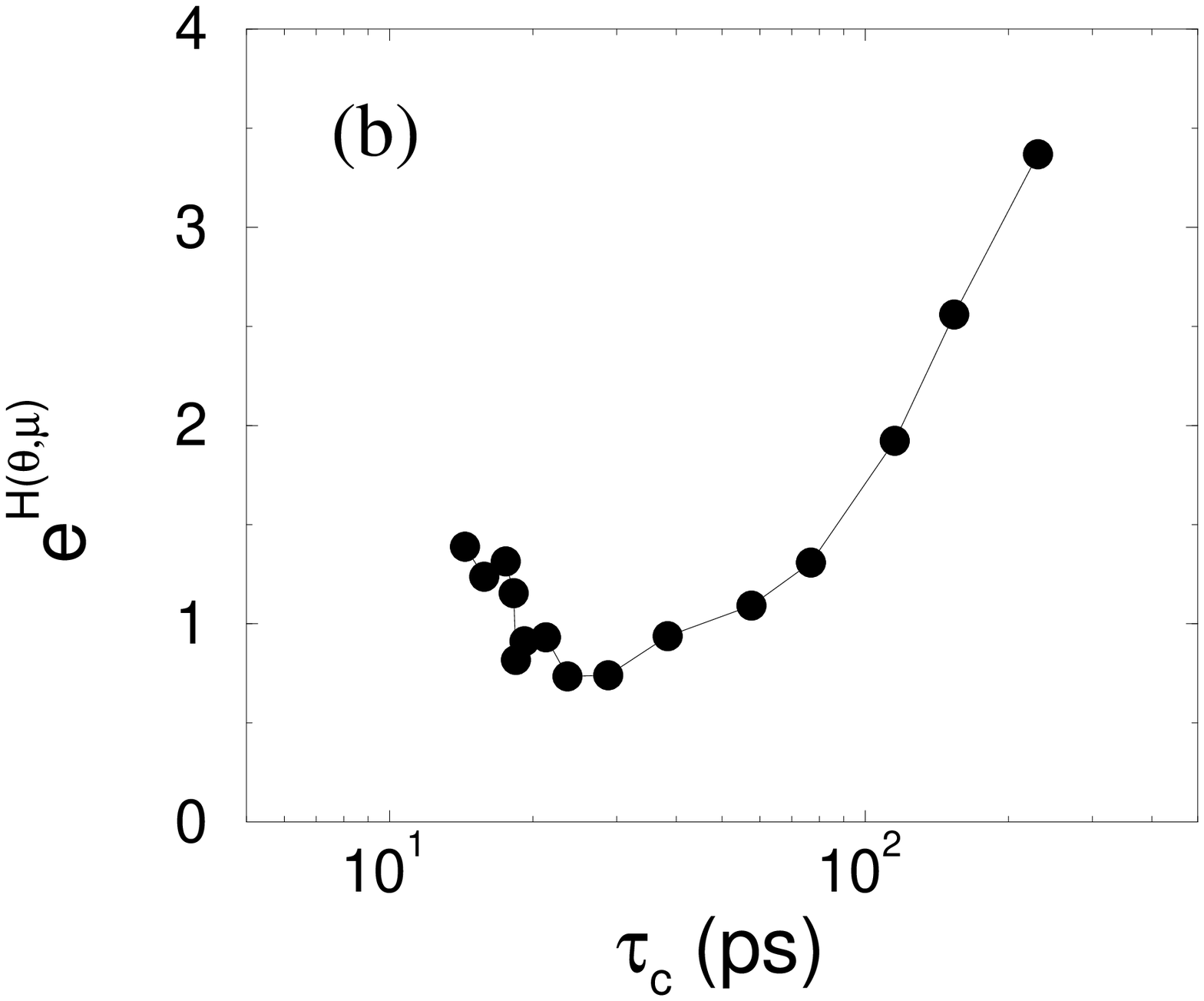,width=4cm}
}
\caption{
Statistical characterization of the noise-correlation coherence resonance.
(a) Standard deviations $R_\theta$ (full circles) and $R_\mu$ (empty squares),
and (b) joint entropy $H(\theta,\mu)$ as a function of the noise correlation
time. The noise amplitude is fixed to $\sigma=0.079$. Other
parameters are those of Fig. \protect\ref{fig:ev-tcorr}.
}
\label{fig:R-ent-tcorr}
\end{figure}

In conclusion, we have shown the existence of double coherence resonance in
an excitable optical system (a semiconductor laser with optical feedback)
driven by an external correlated noise. The
assumption of a non-delta temporal correlation of the noise is fully
meaningful, given the fast characteristic time scale
of the deterministic dynamics of
the system. The pulsed response of the laser, which takes the form of
intensity drop-outs, exhibits a maximal coherence for optimal values of
{\em both} the amplitude and correlation time of the external noise.
These results agree satisfactorily with previous experimental observations.
Other recent investigations have analyzed the influence of correlated noise
in model systems exhibiting coherence resonance \cite{casado}, but in those
cases the corresponding deterministic system lacked a natural second time
scale (different from the optimal pulse separation) which could account for
the resonance with respect to the noise correlation time. In the present
case, however, such a time scale does exist naturally in the system.

We acknowledge financial support from MCyT (Spain), under projects
BFM2000-1108 and BFM2000-0624, and from DGES (Spain), under projects
PB98-0935 and PB97-0141.

}
\end{multicols}

\end{document}